\documentclass[%
 reprint,
 superscriptaddress,
 tightenlines,
 aps,
pra,
 floatfix,
]{revtex4-2}

\usepackage{jabbrv}

\usepackage[american]{babel} 
\usepackage{bm,graphicx,amsmath}
\usepackage[separate-uncertainty=true]{siunitx}
\usepackage{placeins}
\usepackage{amssymb}
\usepackage{amsmath}
\usepackage{epsfig}
\usepackage{nicefrac}
\usepackage{multirow}
\usepackage{mathtools}
\usepackage{lineno}

\usepackage{graphicx} 
\usepackage{braket} 
\usepackage[version=4]{mhchem}
\usepackage[colorlinks=true,linkcolor=blue,citecolor=blue,urlcolor=blue]{hyperref}

\DeclareMathOperator*{\argmin}{arg\,min}
\DeclareMathOperator{\sinc}{sinc}

\DeclareSIUnit\sample{Sa}
\DeclareSIUnit\pixel{px}
\DeclareSIUnit\permille{\text{\textperthousand}}
\DeclareSIUnit{\belmilliwatt}{Bm}
\DeclareSIUnit{\dBm}{\deci\belmilliwatt}

\newcommand*{\secref}[2][]{%
	\hyperref[{sec:#2}]{%
		Sec.~\ref*{sec:#2}%
		\ifx\\#1\\%
		\else
		(#1)%
		\fi
	}%
}

\newcommand*{\sectionref}[2][]{%
	\hyperref[{sec:#2}]{%
		Section~\ref*{sec:#2}%
		\ifx\\#1\\%
		\else
		(#1)%
		\fi
	}%
}

\newcommand*{\figref}[2][]{%
	\hyperref[{fig:#2}]{%
		Fig.~\ref*{fig:#2}%
		\ifx\\#1\\%
		\else
		(#1)%
		\fi
	}%
}

\newcommand*{\eqqref}[2][]{%
	\hyperref[{eq:#2}]{%
		Eq.~(\ref*{eq:#2})%
		\ifx\\#1\\%
		\else
		(#1)%
		\fi
	}%
}

\newcommand*{\figureref}[2][]{%
	\hyperref[{fig:#2}]{%
		Figure~\ref*{fig:#2}%
		\ifx\\#1\\%
		\else
		(#1)%
		\fi
	}%
}

\newcommand*{\figuresref}[2][]{%
	\hyperref[{fig:#2}]{%
		Figures~\ref*{fig:#2}%
		\ifx\\#1\\%
		\else
		(#1)%
		\fi
	}%
}

\begin{document}
\title{Model-Based Real-Time Synthesis of Acousto-Optically Generated Laser-Beam Patterns and Tweezer Arrays}
\author{Marcel Mittenbühler}
\affiliation{Technische Universit\"at Darmstadt, Institut f\"ur Angewandte Physik, Schlossgartenstra\ss e 7, 64289 Darmstadt, Germany}
\author{Lukas Sturm}
\affiliation{Technische Universit\"at Darmstadt, Institut f\"ur Angewandte Physik, Schlossgartenstra\ss e 7, 64289 Darmstadt, Germany}
\author{Malte Schlosser}
\affiliation{Technische Universit\"at Darmstadt, Institut f\"ur Angewandte Physik, Schlossgartenstra\ss e 7, 64289 Darmstadt, Germany}
\author{Gerhard Birkl}
\email[]{For correspondence: apqpub@physik.tu-darmstadt.de}
\homepage[]{https://www.iap.tu-darmstadt.de/apq}
\affiliation{Technische Universit\"at Darmstadt, Institut f\"ur Angewandte Physik, Schlossgartenstra\ss e 7, 64289 Darmstadt, Germany}
\affiliation{Helmholtz Forschungsakademie Hessen f\"ur FAIR (HFHF), GSI Helmholtzzentrum für Schwerionenforschung, 64291 Darmstadt}
\date{\today}

\begin{abstract}
Acousto-optic deflectors (AOD) enable spatiotemporal control of laser beams through diffraction at an ultrasonic grating that is controllable by radio-frequency (rf) waveforms. These devices are a widely used tool for high-bandwidth random-access scanning applications, such as optical tweezers in quantum technology. A single AOD can generate multiple optical tweezers by multitone rf input in one dimension. Two-dimensional (2D) patterns can be realized with two perpendicularly oriented AODs.
As the acousto-optical response depends nonlinearly on the applied frequency components, phases, and amplitudes, and in addition experiences dimensional coupling in 2D setups, intensity regulation becomes a unique challenge. Guided by coupled-wave theory and experimental observations, we derive a compute-efficient model which we implement on a graphics processing unit. Only one-time sampling of single-tone laser-power calibration is needed for model parameter determination, allowing for straight-forward integration into optical instruments. 
We implement and experimentally validate an open-loop diffraction efficiency control system that enables programmable 2D multibeam trajectories with intensity control applied at every time step during digital signal generation, overcoming the limited flexibility, pattern-size constraints, and bandwidth limitations of methods using precalculation and precalibration of a predefined pattern set or closed-loop feedback. The system is capable of stable real-time waveform streaming of arrays with up to 50$\times$50 tweezers with minimal time resolution of \SI{1.4}{\nano\second} (\SI{700}{MS/s}) and a peak latency below \SI{257}{\micro\second} for execution of newly requested patterns.
Reactive, real-time 2D multibeam laser patterning and scanning with strict intensity matching will substantially benefit parallelization and increasing data rates in materials processing, microscopy, and applications of optical tweezers, where the presented work is of immediate relevance for time-critical conditional operations with minimal latency.\\ \textcolor{blue}{Published as Phys. Rev. Appl. $\bf{24}$, 064046 (2025) - DOI: 10.1103/d3tx-3tg8}
\end{abstract}

\maketitle

\section{Introduction}
\begin{figure}[t]
	\includegraphics{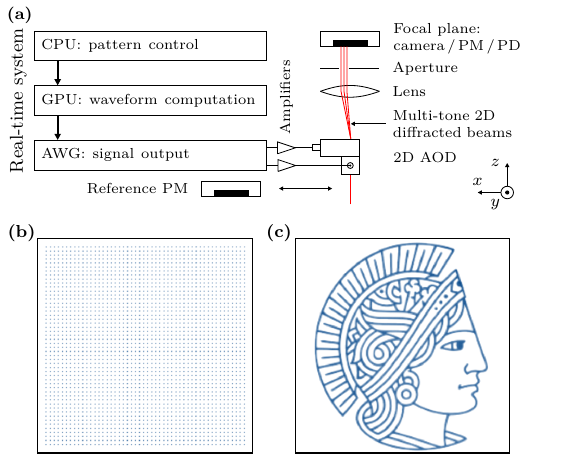}
	\caption{
Real-time synthesis of AOD-generated 2D beam patterns and laser spot arrays. 
(a) Schematic illustration of digital signal synthesis and optical setup (PM, laser power meter; PD, photodiode). The second dimension of the beam pattern is omitted for clarity.
(b) Camera image (\SI{3.1}{\milli\meter}$\times$\SI{3.1}{\milli\meter}) of an intensity-equalized array of 50$\times$50 optical tweezers in the focal plane. For a targeted integral diffraction efficiency of \SI{50}{\percent} an efficiency of \SI{49.5 \pm 2.3}{\percent} is achieved. The frequency span is \SI{36.75}{\mega\hertz} per dimension with a spacing of \SI{0.75}{\mega\hertz} between tones.
(c) Two-dimensional spatiotemporal control was used to imprint the logo of TU Darmstadt depicting the Greek goddess Athene.}
	\label{fig:f0}
\end{figure}

Advanced methods for spatiotemporal control of laser beams are driving innovation in a wide range of applications from quantum technology and advanced manufacturing to biology~\cite{Pinheiro2024,Skliutas2021,Arnoux2022,Bayguinov2018,Elliott2020,Duocastella2021,Mahecic2020,Ashkin1997,Gieseler2021,Moffitt2008,Ricci2024,Ballestero2021,Yan2023,Siegel2025,Liu2022,Manshina2024,Kaufman2021,Cheng2023,Graham2023}. Among the physical principles utilized for dynamic light modulation, the electro-optic and acousto-optic effects facilitate solid-state devices with no moving parts. Compared to inertia-limited spatial light modulators and mirror-based scanners, these optical modulators excel in high-bandwidth applications~\cite{Roemer2014}.
On the basis of the acousto-optic effect (i.e., the diffraction of a laser beam by an ultrasound acoustic wave), optical parameters of the diffracted light can be accessed through the applied waveform of the acoustic wave in the radio-frequency (rf) domain. This allows for precise control of phase, frequency, amplitude, and direction of propagation of the diffracted light. Optimized designs accommodate the basic constraints of phase matching (i.e.~momentum
conservation) for various applications~\cite{Saleh,Chang,Duocastella2021}, combining intensity modulation with beneficial characteristics for rf modulation, tunable wavelength filtering, and beam deflection.\\
In the work presented in this paper, we focus on two-dimensional (2D) intensity and position control, employing two acousto-optical deflectors (AODs) in a mutually perpendicular configuration.
Our methods can be adapted to other acousto-optical designs and applications.
Unlike electro-optic beam scanners, AODs enable parallelization by applying multifrequency waveforms (i.e., multitone patterns of acoustic waves) that generate multiple diffracted beams with a single device. For each beam, the angle of deflection is linearly linked to the specific rf frequency, giving access to straightforward position control.
The ratio of light intensities of diffracted and incident beams, in general, is nonlinearly dependent on the rf power spectrum, relative phases of the individual tones, and geometrical effects. In consequence, intensity regulation presents a considerable challenge, particularly for multitone operation. Practical realizations have resorted to calibration of previously known patterns, which carries at minimum a quadratic overhead with the number of tones, or slow iterative closed-loop feedback, significantly reducing the usable bandwidth.\\
We have developed and implemented a method for the spatiotemporal control of single- and multibeam patterns that harnesses the full acousto-optical bandwidth.
The control system presented mitigates nonlinearities of the acousto-optical response in real time for strict intensity matching in user-defined beam patterns and vector trajectories of laser spots in a 2D plane (\figref{f0}).
For this purpose, we derived a model of the frequency-dependent transfer function that maps the nonuniform diffraction efficiency, the interference of frequency components in multitone patterns, and the implications of using two AODs in series.
Model parameters are independent of the number of rf tones and can be determined via single-tone calibration measurements.
We implement a data-efficient tuning procedure based on model assumptions, that is applied at every time step during digital signal generation using a central processing unit (CPU) as pattern controller, a graphics processing unit (GPU) for waveform computation, and an arbitrary waveform generator (AWG) for signal output (\figref[a]{f0}).
Our open-loop approach is valid for arbitrary configurations and highly scalable.
\\
\hyperref[{fig:f0}]{Figures \ref*{fig:f0}(b) and \ref*{fig:f0}(c)} depict exemplary realizations of 2D multibeam control. In \figref[b]{f0} a uniform 50$\times$50 array of regularly spaced focused Gaussian beams is shown, while the image in \figref[c]{f0} was created by rastering 184 lines of individual one-dimensional (1D) horizontal multibeam patterns for which we implemented an open-loop version of the composite beam method for beam shaping~\cite{Trypogeorgos2013}.
The system presented is particularly suited for improving time-critical random-access scanning applications \cite{Roemer2014,Picard2025,GuoZ2025,Lu2025}. This includes increasing data rates in materials processing by direct laser writing for electronic microfabrication~\cite{Pinheiro2024} and nanoscale additive manufacturing by multiphoton lithography~\cite{Skliutas2021,Arnoux2022}, as well as in advanced imaging techniques of laser-scanning microscopy~\cite{Bayguinov2018,Elliott2020,Duocastella2021} and superresolution imaging~\cite{Mahecic2020}.
Another prominent example is resolved laser addressing for the precise manipulation of individual atomic, molecular, and biological systems with so-called optical tweezers~\cite{Ashkin1997,Gieseler2021}, enabling significant methods in cytology~\cite{Moffitt2008,Ricci2024}, sensing~\cite{Ballestero2021,Yan2023,Siegel2025}, chemistry~\cite{Liu2022,Manshina2024}, and quantum technology~\cite{Kaufman2021,Cheng2023,Graham2023}.\\
In our research work, we target quantum science with atomic and molecular systems in optical tweezer arrays~\cite{Pause2024,Schaffner2024,Schlosser2020}. We use this application to describe and assess our method. In this field~\cite{Kaufman2021}, AOD-generated movable optical tweezers have become the \textit{de facto} standard for realizing transport operations~\cite{Hwang2023,Holland2023,Bluvstein2024,Pause2024,Norcia2024,Gyger2024,Picard2024a,Manetsch2025,Hwang2024,Cicali2025} and laser addressing of quantum systems in a site-selective fashion~\cite{Kim2008,Barnes2022,Yan2022,Bluvstein2024,Picard2024a,Radnaev2024,Yang2024}.
These techniques are essential for the formation of individual molecules~\cite{Liu2022} and the assembly of long-range interacting quantum systems~\cite{Defenu2023} of Rydberg atoms~\cite{Morgado2021} or polar molecules~\cite{cornish2024}, continuous operation of the created quantum arrays~\cite{Pause2023,Norcia2024,Gyger2024,Chiu2025}, and the ability to perform algorithms and implement error correction in neutral-atom quantum processors~\cite{Bluvstein2024,Reichardt2024,Bedalov2024,Rodriguez2024}.
Future progress will significantly benefit from parallelization by flexible multitone operation of AODs with reliable diffraction efficiency.
This is exemplified by the rearrangement of quantum systems for the formation of defect-free arrays, where the number of transport operations shows polynomial scaling with increasing array size. The polynomial degree is above linear for sequential single-tweezer operations~\cite{Schymik2020}. Optimized multitweezer algorithms potentially achieve below square-root scaling, but any practical advantage has been limited by inhomogeneity effects when using multitone waveforms~\cite{Tian2023}.\\
A second parameter of severe practical importance, in particular for conditional mid-circuit operations~\cite{Norcia2023,Bluvstein2024}, is the latency between request and execution of a transport or addressing operation.
Although a hard-wired implementation on a field programmable gate array (FPGA) potentially can achieve superior performance for a fixed algorithm~\cite{Wang2023,GuoX2025}, the accessibility and flexibility of software-based control, scalable GPU compute power, and straightforward AWG-based output have proven to be essential for the presented work.
In contrast to previous approaches on a similar platform~\cite{Tu2024,Dadpour2025}, which inevitably add output delay at the millisecond scale by processing sequences of entire waveforms, our method uses a streaming mode of operation that requires the waveform to be processed continuously in small blocks only. Depending on the size of the block, short latencies of less than \SI{257}{\micro\second} are realized. 
To fulfill these strict real-time requirements on signal generation, we put a highly optimized fixed-core software system into operation. Moreover, the modular programming approach enables the extension to additional applications in areas other than qubit transport by loading additional modules to serve the requirements of specific tasks.\\
The paper is organized as follows. \sectionref{II} summarizes the theory of acousto-optics for the generation of dynamic 2D arrays. \sectionref{III} describes the model for intensity adjustment via 2D decomposition and single-axis diffraction efficiency control, as well as details on waveform streaming performance and the real-time signal generation framework. In \secref{IV} the test apparatus is explained and measurements of the single- and multitone efficiency control are presented.

\section{Theory of acousto-optics for dynamic tweezer arrays}
\label{sec:II}
The interaction between an incident light field and the acoustic wave within the crystal of an AOD can be understood as a scattering process. The acoustic wave changes the refractive index, creating a periodic grating that diffracts the incoming light wave~\cite{Saleh,Chang}. Assuming plane waves and a fixed optical wavelength, modeling the diffraction efficiency in the first diffraction order needs mainly to consider two dependencies. First, maximum diffraction occurs as a result of wave-vector matching within the crystal and therefore is a function of the acoustic waveform~\cite{korpel1981acousto,Chang,Antonov2019}.
Second, coupled-wave theory provides a foundation for the dependence of the diffraction efficiency $\eta(\nu)$ on the power $P_{\rm a}(\nu)$ of the acoustic wave with frequency $\nu$ and the properties of the AOD crystal parameterized in the acousto-optical figure of merit $\eta_{\rm c}$. Under the above assumptions, $\eta_{\rm c}$ is constant for a particular device. The single-tone diffraction efficiency $\eta(\nu)$ is given by Ref.~\cite{Chang} using the normalized sinc function:
\begin{equation}
    \eta(\nu) = \eta_\textnormal{c} P_\textnormal{a}(\nu) \sinc^2\left(\frac{1}{2\pi}\sqrt{4\eta_\textnormal{c} P_\textnormal{a}(\nu) + (\Delta k(\nu) L)^2}\right)
    \label{eq:k-mismatch-model}
\end{equation}
with crystal length $L$ and wave vector mismatch $\Delta k(\nu) = |\vec{k} + \vec{q}(\nu) - \vec{k_\textnormal{d}}(\nu)|$ of the incident light wave $\vec{k}$, the acoustic wave $\vec{q}(\nu)$, and the diffracted light wave $\vec{k_\textnormal{d}}(\nu)$.\\
Typically, AODs are operated in a power regime up to the first maximum of $\eta(\nu)$. In negligence of the oscillatory behavior, this maximum is commonly termed saturation.
We approximate $\eta(\nu)$ by introducing the experimentally accessible quantities of peak diffraction efficiency $\alpha(\nu)$, rf power $P(\nu)$, and rf saturation power $\beta(\nu)$ as 
\begin{equation}
    \eta(\nu) \simeq \alpha(\nu) \sin^2 \left ( \frac{\pi}{2} \sqrt{\frac{P(\nu)}{\beta(\nu)}} \right ).
    \label{eq:k-mismatch-approx}
\end{equation}
The error incurred by this approximation up to saturation, where $P(\nu)=\beta(\nu)$ yields $\eta(\nu)\simeq\alpha(\nu)$, is below \SI{0.03}{\percent}. The dimensionless fraction $P(\nu)/\beta(\nu)$, describing the power scaling, admits to rescale the expression to any experimentally accessible power measure, for example, the output power of the rf signal generator or the digital power level set to drive it. \\
For waveform synthesis, it is central to determine the value of $P(\nu)$ required to achieve a targeted diffraction efficiency. The approximate model can be inverted efficiently. Moreover, using coupled-wave theory, a simple analytical solution for multitone acoustic waveforms (i.e., operation with multifrequency rf signals) can be derived. 
Let $\eta(\nu) \in [0, 1]^N$ be a list of $N$ diffraction efficiencies at respective frequencies $\nu \in \mathbb{R}^N$ and rf powers $P(\nu) \in \mathbb{R}^N$. Assuming no coupling between diffracted components and operation close to the Bragg regime ($\Delta k \approx 0 $), we find
\begin{equation}
    \eta_i := \eta(\nu_i) = \frac{ \alpha \left (\nu_i \right ) P(\nu_i) }{\beta \left (\nu_i \right) \sum_{l=1}^N \frac{P(\nu_l)}{\beta \left ( \nu_l \right)}} \sin^2 \left ( \frac{\pi}{2} \sqrt{ \sum_{l=1}^N \frac{P(\nu_l)}{\beta \left ( \nu_l \right)} } \right ).
    \label{eq:1d-model}
\end{equation}
Taking into account normalization of all $P(\nu_i)$ to $\beta \left (\nu_i \right)$,  the global diffraction efficiency depends only on the total rf power. In addition, the global diffraction efficiency is divided among all tones $\nu_i$ by their normalized fraction of the total rf power.\\
Next, we consider two AODs in a sequential, mutually perpendicular configuration for 2D deflection, as depicted in \figref[a]{f0}. The angle of the incident beam on the second AOD is altered by the deflection of the first, depending on the frequency components of the applied rf signal. Thus, the 2D diffraction efficiency must account for the deflection in both AODs, which gives rise to a reduction of the peak diffraction efficiency of the 2D system due to the induced wave vector mismatch.
We model the 2D diffraction efficiency $\eta^{(\textnormal{xy})}_{i,j}$ by single-axis efficiencies $\eta^{(\textnormal{x})}_i$, $\eta^{(\textnormal{y})}_j$ and an interaction term $\Xi_{i,j}$ that expresses the frequency-dependent efficiency loss:
\begin{equation}
    \eta^{(\textnormal{xy})}_{i,j} = \eta^{(\textnormal{x})}_i \eta^{(\textnormal{y})}_j \Xi_{i,j}.
    \label{eq:eta-xy-model}
\end{equation}
For deflection with $N$ tones $\nu_i^{(\textnormal{x})}$ in $x$ and $M$ tones $\nu_j^{(\textnormal{y})}$ in $y$, $\Xi \in [0, 1]^{N \times M}$.
It is considered an error term which leads to a deviation from two ideal single-axis systems. To minimize the error between the measurement and our model, we analyzed the dominant terms in a Taylor series, finding empirically that the residuals in $\eta^{(\textnormal{xy})}_{i,j}$ can be minimized assuming
\begin{equation}
    \Xi_{i,j} = \left (1-\xi_0 \left(\nu^{(\textnormal{x})}_i-\xi_\textnormal{x}\right)^2 \left(\nu^{(\textnormal{y})}_j-\xi_\textnormal{y} \right)^2 \right)^2,
    \label{eq:xi-def}
\end{equation}
with interaction parameters $\xi_\textnormal{x}$, $\xi_\textnormal{y}$, and $\xi_0$. The parameters $\xi_\textnormal{x}$ and $\xi_\textnormal{y}$ are frequencies related to the geometry of the optical setup, while $\xi_0$ quantifies the interaction strength. For different applications and AOD devices, a revised form of $\Xi$ may be chosen for minimizing the residuals.

\section{Real-time modeling and signal generation}
\label{sec:III}

In a typical application, a target diffraction efficiency pattern $\eta^{(\textnormal{xy})}_\textnormal{T} \in [0, 1]^{\rm{N\times M}}$ should be generated with frequencies $\nu^{(\textnormal{x})} \in \mathbb{R}^N, \nu^{(\textnormal{y})}\in \mathbb{R}^M$ defining the spatial structure of the pattern. 
The goal of the method presented is the computation of the required rf power values for all frequency components that minimize the error between the target efficiency pattern and the measured diffraction efficiencies. 
As a simplification, we assume that $\eta^{(\textnormal{xy})}_\textnormal{T}$ can be decomposed into an outer product defined by the single-axis target diffraction efficiencies $\eta^{(\textnormal{x})}_\textnormal{T} \in [0, 1]^N, \eta^{(\textnormal{y})}_\textnormal{T} \in [0, 1]^M$. Due to the interaction term, it is impossible to achieve the exact pattern in a 2D array if $N > 1 \land M > 1$; the interaction can be exactly compensated for a single beam and a line of beams only.
However, it is possible to compute sets of adjusted diffraction efficiencies $\eta^{(\textnormal{x})}_i$ and $\eta^{(\textnormal{y})}_j$ to minimize the error caused by $\Xi$.
Consequently, our approach splits the model-based diffraction control into two parts. First, the adjusted $\eta^{(\textnormal{x})}_i$ and $\eta^{(\textnormal{y})}_j$ are calculated to minimize the error in $\eta^{(\textnormal{xy})}_{i,j}$ compared to the targeted 2D diffraction efficiency $\eta^{(\textnormal{xy})}_\textnormal{T}$. Second, the 1D diffraction efficiency model is used to compute the rf power values needed for achieving the adjusted diffraction efficiencies for each axis.
\subsection{Two-dimensional decomposition}
The actual diffraction efficiency given in Eq. \eqref{eq:eta-xy-model} contains the interaction term that restricts an exact decomposition when controlling multiple tones in both dimensions.
The goal is to minimize the error in the achieved diffraction efficiency caused by $\Xi_{i,j}$. Therefore, adjusted values for the 1D diffraction efficiency $\eta^{(\textnormal{x})}_i, \eta^{(\textnormal{y})}_j$ need to be found, as a minimization problem with
\begin{equation}
    \argmin_{\eta^{(\textnormal{x})}_i, \eta^{(\textnormal{y})}_j} \sum_{i=1}^N \sum_{j=1}^M \left ( \eta^{(\textnormal{x})}_{\textnormal{T},i} \eta^{(\textnormal{y})}_{\textnormal{T},j} - \eta^{(\textnormal{x})}_{i} \eta^{(\textnormal{y})}_{j}\Xi_{i,j}\right )^2.
\end{equation}
An approximate solution given by a 2D decomposition can be computed by
\begin{align}
    \eta^{(\textnormal{x})}_i &= \frac{\eta^{(\textnormal{x})}_{\textnormal{T},i}}{ \sqrt{c_0} c_\textnormal{x} \sum_{j=1}^M \eta^{(\textnormal{y})}_{\textnormal{T},j} \Xi_{i,j}},\label{eqn:compterm1}\\
    \eta^{(\textnormal{y})}_j &= \frac{\eta^{(\textnormal{y})}_{\textnormal{T},j}} {\sqrt{c_0} c_\textnormal{y} \sum_{i=1}^N \eta^{(\textnormal{x})}_{\textnormal{T},i} \Xi_{i,j}},\label{eqn:compterm2}\\
    c_0 &= \left[ \sum_{i=1}^N \eta^{(\textnormal{x})}_{\textnormal{T},i} \sum_{j=1}^M\eta^{(\textnormal{y})}_{\textnormal{T},j} \Xi_{i,j} \right]^{-1}.
    \label{eqn:normterm1}
\end{align}
The solution is not uniquely specified as it can be altered without consequences by a constant ratio of the efficiencies $\nicefrac{c_\textnormal{x}}{c_\textnormal{y}}$ with $c_\textnormal{x} c_\textnormal{y} = 1$. This degree of freedom mirrors the experimental implementation since diffraction efficiencies are measured after both AODs only. Finding the exact ratio can be challenging as this would require a secondary optimization goal to introduce a constraint, for example, measuring the diffraction efficiency for both AODs independently. It is a strength of our approach that these additional measurements are not required as the ratio can be chosen to distribute the mean deviations from the target diffraction efficiency approximately equally between both axes, by setting
\begin{equation}
    c_\textnormal{x} = \frac{1}{c_\textnormal{y}} = \sqrt{\frac{\sum_{i=1}^N \sum_{j=1}^M \eta^{(\textnormal{y})}_{\textnormal{T},j} \Xi_{i,j}}{\sum_{i=1}^N \sum_{j=1}^M \eta^{(\textnormal{x})}_{\textnormal{T},i} \Xi_{i,j}}}.
    \label{eqn:normterm2}
\end{equation}
In general, computation of Eqs. \eqref{eqn:compterm1} and \eqref{eqn:compterm2} would require a summation over the target diffraction efficiency and the interaction term $\Xi$ for each tone, resulting in quadratic scaling in the number of frequency components. Moreover, Eqs. \eqref{eqn:normterm1} and \eqref{eqn:normterm2} would require a double summation, resulting in quadratic scaling of the computation time concerning the number of frequency components.
The form of $\Xi$ in Eq. \eqref{eq:xi-def} on the other hand, allows a decomposition such that a summation can be rewritten as
\begin{equation}
    \begin{aligned} \sum^M_{j=1} \eta^{(\textnormal{y})}_{\textnormal{T},j} \Xi_{i,j} = a^{(0)}_y 
 &+ \left ( \nu^{(\textnormal{x})}_i - \xi_\textnormal{x} \right )^2 a^{(1)}_y \\ &+
    \left ( \nu^{(\textnormal{x})}_i - \xi_\textnormal{x} \right )^4 a^{(2)}_y,
\end{aligned}
\end{equation}
\begin{align}
    a^{(0)}_y &= \sum^M_{j=1} \eta^{(\textnormal{y})}_{\textnormal{T},j},\\
    a^{(1)}_y &= \sum^M_{j=1} -2 \eta^{(\textnormal{y})}_{\textnormal{T},j} \xi_0 \left ( \nu^{(\textnormal{y})}_j - \xi_\textnormal{y} \right )^2,\\
    a^{(2)}_y &= \sum^M_{j=1} \eta^{(\textnormal{y})}_{\textnormal{T},j} \xi_0^2 \left ( \nu^{(\textnormal{y})}_j - \xi_\textnormal{y} \right )^4.
\end{align}
By performing this decomposition, the y-axis constant factors $a^{(0,1,2)}_y$ are computed just once for each time step, requiring linear time. Computation of Eq. \eqref{eqn:compterm1} requires constant time, so it can be computed for all frequency components in linear time. For the x axis, the constant factors $a^{(0,1,2)}_x$ can be computed analogously. Thus, the normalization factors $c_0$, $c_\textnormal{x}$, $c_\textnormal{y}$ and the adjusted 1D diffraction efficiencies $\eta^{(\textnormal{x})}_i, \eta^{(\textnormal{y})}_j$ can be computed efficiently. In general, such a decomposition can be performed for other forms of $\Xi$ as long as they consist of polynomials.
\subsection{Single-axis diffraction efficiency control}
For each axis, the required adjusted diffraction efficiencies are achieved by tuning the rf power for each tone.
Equation \eqref{eq:1d-model} admits an exact solution for the required rf power of each component:
\begin{equation}
    P(\nu_i) = \mkern-18mu \underbrace{\frac{\eta_i \beta(\nu_i) }{\alpha(\nu_i)}}_{\textnormal{local scaling factor}} \underbrace{ \frac{4}{\pi^2 \sum_{l=1}^N \frac{\eta_l}{\alpha(\nu_l)}} \arcsin^2 \sqrt{\sum_{l=1}^N \frac{\eta_l}{\alpha(\nu_l)}} }_\textnormal{global scaling factor}.
    \label{eq:power1D}
\end{equation}
For the signal computation it is beneficial to initially consider the local power terms only and then rescale the summed signal. This requires a single $\arcsin$ computation and a single iteration over all tones.\\
We found that both the peak diffraction efficiency $\alpha$ and the saturation power $\beta$ can be modeled well by a power series (see \secref{modelparameters}).
To efficiently compute $\alpha$ and $\beta$, the values for the functions are discretized and stored in constant memory (256 points per dimension).\\
For readout, linear interpolation is performed for a smooth transition between discretization points. As the frequency of each tone changes only slowly, it is almost constant over short time scales. Therefore, spatially related memory regions are accessed for computing $\alpha$ and $\beta$, resulting in fast read performance.\\
It is important to note that, if not properly optimized, multitone signals can easily surpass the regime of linear amplification due to constructive interference of frequency components. Depending on the relative phases, different peak-to-mean amplitude ratios are incurred, also called crest factors.
In general, low crest factors are desirable as they result in better signal-to-noise ratios reducing intermodulation~\cite{Endres2016} and may not exceed power specifications of the AOD. Crest factor optimization is beyond the scope of this work as it is a fundamental problem in rf signal engineering~\cite{Carvalho_Jang_2011}. In applications of optical tweezers using static beam patterns known ahead of time, it has proven useful to iteratively optimize the phases~\cite{guillaume1991crest, van1988peak, van1987new} as it significantly improves signal quality.
As it is not possible to compute all possible frequency combinations in advance, a direct method needs to be used for phase optimization ~\cite{kitayoshi1985dsp, newman19651, schroeder1970synthesis, Narahashi1995, shibasaki2020analysis}. For the reactive open-loop control system presented, we use the precalculation described in Ref.~\cite{Narahashi1995} commonly known as the Narahashi phases, as the computational effort is manageable, but provides a better crest factor reduction compared to other direct methods.
\subsection{Real-time system}
The system used in this work (\figref[a, left]{f0}) consists of a controller on the CPU that manages which 2D beam patterns should be generated, the signal computation on the GPU, and finally the signal output using an AWG. The data transfer between GPU and AWG is performed via remote direct memory access (RDMA) to minimize latency and CPU load. 
The AWG runner is highly isolated due to the strict real-time requirements of the AWG in streaming mode. Timing jitter above \SI{10}{\micro\second} can result in a buffer underflow, crashing the AWG.
The RDMA is performed in blocks that get transferred at a time before they can be output. 
On the GPU, at least two data blocks should be allocated such that while one is transferred to the AWG, the other can be written with the subsequent waveform. 
The buffer sizes directly impact the achievable latency of the system, as in the worst case all buffers need to be output before a newly requested waveform gets generated. Therefore, the buffer size should be chosen as small as possible while keeping the system stable.
The time bin defined by buffer size and sample rate must suffice for all communication between CPU, GPU, and AWG, as well as the computation of a new block and the data transfer. To avoid buffer underflow, a real-time system is crucial.
As only one short part of the waveform is computed at a time, the reference points for GPU computation are regularly updated by the CPU with each requested waveform block. This approach prevents loss of precision caused by the use of single-precision floating-point numbers on the GPU. 
\subsection{Waveform streaming performance}
\label{sec:streaming}
The hardware for this real-time system consists of an AMD Ryzen Threadripper PRO 5955WX CPU, an Nvidia Ada 4500 graphics card containing the GPU for waveform computation, and an additional graphics card for rendering tasks. Moreover, the system houses the Spectrum M4i.6631-x8 AWG (two channels, 16 bit, \SI{1.25}{GS/s}). All real-time devices are attached to NUMA node 1 which is isolated from other processes, with the GPU and AWG being within the same root complex for ideal data transfer conditions.
We do not reach the maximum output sample rate of the AWG, as the PCI Express implementation (Gen2 x8) limits continuous data streaming to \SI{2.8}{GB/s}, resulting in \SI{700}{MS/s} per channel for our 2D system. Hence, at each channel the signal is output with a time resolution of \SI{1.4}{\nano\second}.\\
The user defines the requested waveform through instruction sets that comprise initial conditions for frequency, phase, and diffraction efficiency of all tones, as well as their derivatives defining waveform evolution for a specified time step.
By stitching multiple sets, dynamic pattern sequences defined by flexible multitone parameter trajectories can be programmed. The queue is managed by the CPU and continuously transmitted to the GPU.
Subsequent queue entries are output at the full time resolution of \SI{1.4}{\nano\second} without additional latency between arbitrary tone patterns.\\
Assuming an empty queue, the peak latency between requesting a new pattern and the output of this pattern as analog waveform is defined by the signal buffer sizes on the GPU and AWG. Minimization of the transfer size to \SI{256}{\kilo\byte} for the AWG with double buffering on the GPU and a single buffer on the AWG results in a peak latency of \SI{257}{\micro\second} and a minimal latency of \SI{164}{\micro\second}.\\
In this configuration, our software implementation for multitone rf signal processing on the GPU achieves sustained computation of waveforms including model-based efficiency control for arbitrary patterns of up to 50$\times$50 tones.
The system handles static and dynamic patterns identically. Thus, all patterns presented in this paper have been generated in real-time mode. In addition, the computation handles all patterns as if they consist of 50 tones, some of which may have zero amplitude. This approach ensures that the system is stable independently of the pattern. It also simplifies code optimization because an empty pattern has the same computational overhead as a dynamic 50$\times$50 pattern.
The system is stable over multiple days without any buffer underflows. All performance parameters can be further improved with more powerful state-of-the-art hardware.

\section{Experimental Validation}
\label{sec:IV}
We now describe the implementation and experimental evaluation of this system. 
Measured diffraction efficiencies always reflect the combined response of both AODs and are labeled $\eta^{(\textnormal{xy})}_\textnormal{M}$.
Absolute rf power levels are given as root-mean-square (rms) values. 
The AWG output for each channel is amplified by a Minicircuits ZHL-03-5WF+ power amplifier.
We calibrated the AWG's output range such that for an rms rf power of \SI{2}{\watt} of a single-tone signal at a frequency of \SI{101}{\mega\hertz} the AWG power fraction is defined as \num{1}.
For parameter sampling and measurements, we parametrize the applied rf power by the output power fraction of the AWG card that is digitally set in the interval $[0, 1]$.
Unless stated otherwise, uncertainties are given as the 95th percentile (i.e.,~$2\sigma$ deviation) throughout this paper.

\subsection{Optical setup: 2D AOD tweezers}
The optical setup is schematically depicted in \figref[a]{f0}. The AA Opto-Electronics DTSXY-400-800 combines two single-axis AODs that are oriented perpendicularly to deflect along the $y$ and $x$ dimensions (in that order). The AODs are based on a \ce{TeO_2} crystal in shear-mode configuration with an active aperture of \SI{7.5}{\milli\meter}$\times$\SI{7.5}{\milli\meter}. The optical transmission per AOD is \SI{99}{\percent} according to the data sheet.
\\
Light at a wavelength of \SI{800}{\nano \meter} is supplied via a single-mode polarization-maintaining fiber with linear polarization matched to the input polarization axis of the first AOD. 
The laser beam is collimated to a beam waist of \SI{1.7}{\milli\meter} ($1/e^2$~radius) at the position of the AOD. Behind the AOD, a lens with focal length of \SI{63.5}{\milli\meter} 
focuses the diffracted beam such that the angular deflection is turned into a displacement in the back focal plane.
An aperture blocks any order other than the desired first order. The front focal plane of the lens is positioned approximately in the middle of the two AOD crystals. The selected optical parameters, specifically the beam size at the AOD crystals, are predetermined by our targeted application. We expect no performance limitations to arise when using the full available aperture of the AODs. \\
An initial evaluation of the AOD showed that the incident beam and the diffracted first-order beam are collinear at a frequency of \SI{104}{\mega\hertz}.
The minimum of the central dip in diffraction efficiency, typical for wide-angle AODs using birefringent crystals (see \figref[a]{model}), is at $\nu_\textnormal{c}=\SI{101}{\mega\hertz}$. We use  $\nu_\textnormal{c}$ as center frequency for the scanning interval and corresponding measurements in the work presented. The approximate bandwidth of \SI{42}{\mega\hertz} results in a scan angle of \SI{50.2}{\milli\radian} and translates into a 2D scan range of \SI{3.16}{\milli\meter}$\times$\SI{3.16}{\milli\meter} in the back focal plane for the selected focal length. The focused spot waist is \SI{11}{\micro\meter} ($1/e^2$~radius).
For applications that require a modified addressable area, spatial resolution, or working distance, the back focal plane can be relayed using reimaging optics (not shown).
A power meter with a thermopile sensor measures the laser power. The total measurement uncertainty, including calibration uncertainty, repeatability, linearity, and uniformity errors is specified to \SI{\pm 3}{\percent}. For relative measurements, the calibration uncertainty can be neglected. In this case, the resulting uncertainty is \SI{\pm 2.3}{\percent}.
All measurements use random sampling to reduce measurement errors due to thermal drifts that depend on previously measured power levels. When applicable, we performed a baseline measurement after heating the sensor to compensate for this effect. In addition, fits include an offset term.\\
The laser power was permanently stabilized using a photodetector and digital intensity controller~\cite{Preuschoff2020} (not shown). The reference power level was \SI{750}{\milli \watt}, measured directly in front of the AOD by the reference power meter (see \figref[a]{f0}).
The measured diffraction efficiencies $\eta^{(\textnormal{xy})}_\textnormal{M}$ are defined as the ratio of the measured power of the first diffracted order in the back focal plane and the reference power.
Images of the beam pattern in the back focal plane [\figref[b]{f0}, \figref[c]{f0}, and \figref[a]{LightIntermodulation}] were recorded with a CMOS camera (sensor size \SI{5.60}{\milli\meter}$\times$\SI{3.16}{\milli\meter}, pixel size \SI{1.45}{\micro\meter}$\times$\SI{1.45}{\micro\meter}) without additional imaging optics.
For camera measurements, the intensity stabilization was disabled and the input laser power was reduced to avoid saturation of the CMOS sensor.
\subsection{Model parameter determination}
\label{sec:modelparameters}
\begin{figure}[bt!]
	\includegraphics{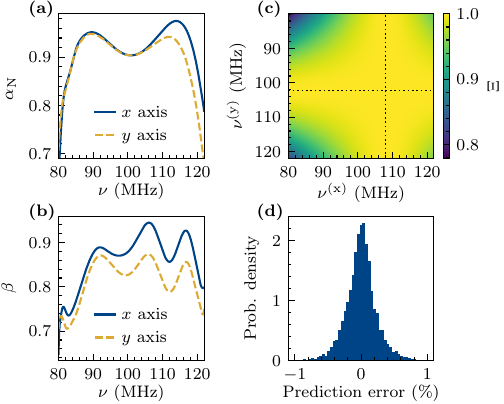}
	\caption{
    Parameter determination for model-based diffraction efficiency control using randomly sampled single-tone data. (a) Peak diffraction efficiency $\alpha_{\text{N}}$ and (b) saturation power $\beta$ in units of the AWG power fraction were modeled for the x-axis AOD (solid blue) and the y-axis AOD (dashed yellow).
    (c) The interaction term $\Xi$ resolves correlations between the two crossed AODs. (d) Histogram of the residual prediction error. The measured data are modeled with an error below \SI{\pm 0.7}{\percent} (2$\sigma$ deviation).
    }
	\label{fig:model}
\end{figure}
The model-based control requires predetermined values of $\alpha$, $\beta$, and the interaction parameters $\xi_0$, $\xi_\textnormal{x}$, and $\xi_\textnormal{y}$. 
For this purpose, the AOD response in terms of $\eta^{(\textnormal{xy})}_\textnormal{M}$ was sampled in the relevant frequency range in a randomized fashion and a least-squares estimator was used to determine the model parameters. 
As per measurement protocol, losses at optical surfaces, apertures, and within optical components are captured in $\eta^{(\textnormal{xy})}_\text{M}$ and thereby incorporated in $\alpha$ and the interaction parameters of the model.\\
We performed sample acquisition using single-tone signals on both AODs with frequencies uniformly drawn from continuous values between \SI{80}{\mega\hertz} and \SI{122}{\mega\hertz} and AWG power fractions also uniformly between \num{0} and \num{1}.
A total of \num{20e3} measurements were collected for sufficient statistics. 
For $\alpha$ and $\beta$ polynomials of different degrees were fitted and rated by the model error on a validation dataset. We settled on polynomials of degree $N=40$ as the error was minimized and no overfitting was observed.
Our approximation of the diffraction efficiency is only valid up to the first diffraction maximum of the AOD (see \secref{II}). Collected samples exceeding saturation adversely impact the model accuracy. For this reason, we first fitted the model with all data points and used the approximate values of $\beta$ to remove samples beyond saturation. Then the fit was repeated for the remaining \num{14e3} samples.\\
It is important to note that during parameter sampling, but also for typical applications of the model-based control, only the product $\alpha^{(\textnormal{x})}\alpha^{(\textnormal{y})}$ needs to be determined. From this combined measurement, the individual peak diffraction efficiency of each axis can be extracted only up to a multiplicative constant. 
We introduce the normalized peak diffraction efficiencies
\begin{align}
    \alpha^{(\textnormal{x})}_\textnormal{N} &= c_{\alpha} \alpha^{(\textnormal{x})} \\
    \alpha^{(\textnormal{y})}_\textnormal{N} &= \frac{1}{c_{\alpha}} \alpha^{(\textnormal{y})} 
\end{align}
with a normalization constant $c_\alpha$ in analogy to the discussion of Eqs. \eqref{eqn:compterm1} - \eqref{eqn:normterm2}.
In practice, our approach uses the identical fraction of the peak diffraction efficiency for both axes when determining the required rf power for a specific requested single-axis diffraction efficiency (Eq. \eqref{eq:power1D}). Therefore, the normalization constant does not influence the model performance. For the complete 2D expression, the normalization cancels in Eq. \eqref{eq:eta-xy-model} due to $c_{\alpha} \times \frac{1}{c_{\alpha}} = 1$.
As such, we fitted the normalized peak diffraction efficiencies $\alpha^{(\textnormal{x})}_\textnormal{N}, \alpha^{(\textnormal{y})}_\textnormal{N}$
instead of
$\alpha^{(\textnormal{x})}, \alpha^{(\textnormal{y})}$ and used their values for the model.
The additional degree of freedom introduced by normalization is eliminated by setting
\begin{equation}
    \alpha^{(\textnormal{x})}_\textnormal{N}(\nu_\textnormal{c}) = \alpha^{(\textnormal{y})}_\textnormal{N}(\nu_\textnormal{c}).
\end{equation}
This choice maximizes the usable scanning range for our AOD model assuming that we request equal diffraction efficiencies per axis. Different normalizations may be beneficial for other devices.\\
\figuresref[a]{model} - \ref{fig:model}\textcolor{blue}{(c)} visualize the final fit parameters. The peak diffraction efficiency $\alpha_\textnormal{N}$ [\figref[a]{model}] expresses the quality of wave vector matching and therefore reflects the expected band shape of the acousto-optic interaction in the case of birefringent diffraction~\cite{Chang}. The saturation power $\beta$ [\figref[b]{model}] shows a more complex behavior. While $\beta$ is expected to decrease with increasing wave vector mismatch, which is the case for the lowest and highest frequencies, additional frequency-dependent transfer functions, such as the frequency response of the power amplifier, the piezoelectric transducer, and the impedance matching circuit within the AOD, strongly influence the obtained shape.
Our approach directly adjusts the digital output power of the AWG card to compensate for frequency-dependent rf power changes.
From the presentation of the interaction term $\Xi$ [\figref[c]{model}] it becomes evident that the geometric factors in the 2D system account for a reduction in diffraction efficiency of the order of \SI{20}{\percent} at the edges of the accessible frequency range.
The extracted value of the interaction strength $\xi_0 = \num{0.94(1)}$. The dashed lines in \figref[c]{model} mark the frequencies for least geometric impairment, $\xi_\textnormal{x} = \SI[separate-uncertainty=false]{108.02(8)}{\mega\hertz}$ and $\xi_\textnormal{y} = \SI[separate-uncertainty=false]{102.20(3)}{\mega\hertz}$.
The deviation from $\nu_\textnormal{c}$ indicates that the alignment, (i.e., the wave vector matching) was better along the y axis than the x axis. This is also visible in the stronger asymmetry observed in $\alpha_\textnormal{N}$ along the x axis in \figref[a]{model}.
The following subsections prove that our model-based control can easily compensate for potential wave vector mismatches.
\figureref[d]{model} shows a histogram of the residual prediction error of the fit model, which is defined as the difference between the measured value $\eta_\textnormal{M}^{(\textnormal{xy})}$ and the model-predicted diffraction efficiency $\eta_{i,j}^{(\textnormal{xy})}$ for each sample.
On basis of the determined model parameters, \SI{95}{\percent} of the residuals fall within \SI{\pm 0.7}{\percent} diffraction efficiency error for the full frequency range (2$\sigma$ deviation).

\subsection{Real-time performance}
\begin{figure}[bt!]
	\includegraphics[width=\linewidth]{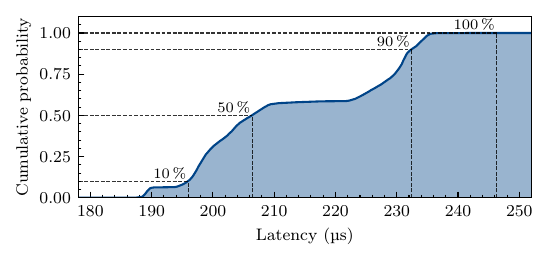}
	\caption{Measured latency between submission of a pattern to an empty data queue (t = 0) and detection of the requested light pattern on a photodiode ($10^4$ samples). A peak latency below \SI{247}{\micro\second} is measured.}
	\label{fig:latency}
\end{figure}
All beam patterns presented in this paper are generated in real-time operation. To validate the latency specification derived from digital data flow analysis in Sec.~\ref{sec:streaming} in the optical signal, the actual latency between submitting a new pattern and recording the resulting light signal was measured. For this, the pattern queue was emptied, so that a zero-volt signal level is output to the AOD, blanking the deflected beam pattern. Then the control program emits an output trigger and directly thereafter submits a pattern to the queue that generates a beam pattern illuminating a fast photodiode. An oscilloscope measures the delay between the control trigger and a light signal on the photodiode for $10^4$ samples. 
The cumulative distribution of latencies in \figureref{latency} reveals that the median latency is below \SI{207}{\micro\second} and peak latency below \SI{247}{\micro\second}. The measured upper bound is even lower than the specified peak latency confirming real-time operation with peak latency of \SI{257}{\micro\second}. 
\subsection{Single-Tone Efficiency Control}
\begin{figure}[bt!]
    \centering
    \includegraphics[width=\linewidth]{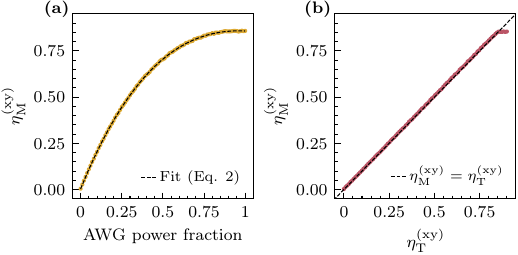}
    \caption{
   Measured diffraction efficiency with and without efficiency control. A single tone is applied to both axes. (a) Without efficiency control, a constant AWG power fraction of \num{0.9} is set for the y axis while sweeping the x-axis AWG power fraction between \num{0} and \num{1}. (b) Efficiency control set to \num{0.9} diffraction efficiency for the y axis while sampling the x-axis diffraction efficiency from \num{0} to \num{1} (i.e.~$\eta_\textnormal{T}^{(\textnormal{xy})}\in[0, 0.9]$) results in a linear behavior up to saturation.
    }
    \label{fig:linearity}
\end{figure}
In a second set of measurements, we validated the capabilities of the diffraction efficiency control system for single-tone applications. For comparison, we performed characteristic measurements of the AOD response as a function of the applied rf power without diffraction efficiency control.\\
\figureref[]{linearity} depicts the measured diffraction efficiency when driving the y axis with a single tone at $\nu_\textnormal{c}$ and the x axis with a single tone at \SI{116}{\mega\hertz}.
The behavior of $\eta^{(\textnormal{xy})}_\textnormal{M}$ as a function of the x-axis AWG power fraction is shown in \figref[a]{linearity} for disabled efficiency control. The y-axis AOD was driven at a constant AWG power fraction of \num{0.9} and the x-axis power fraction was sampled between \num{0} and \num{1}. 
The nonlinear AOD response is well described by Eq. \eqref{eq:k-mismatch-approx} (dashed line),
where all constant losses, including the effect of the static y-axis AOD, are incorporated into $\alpha_\textnormal{N}$.
\\
In comparison, we used our model-based control system to set the target diffraction efficiency $n_\textnormal{T}^{(\textnormal{y})}=\num{0.9}$, while sampling $n_\textnormal{T}^{(\textnormal{x})}\in[0, 1]$. This defines the measurement range $\eta_\textnormal{T}^{(\textnormal{xy})}\in[0, 0.9]$ depicted in \figref[b]{linearity}.
The linear scaling of $\eta^{(\textnormal{xy})}_\textnormal{M}$ as a function of $\eta^{(\textnormal{xy})}_\textnormal{T}$ under efficiency control is evident and well described by the identity relation (dashed line) up to the point where the AOD reaches its maximum efficiency and the response saturates.
\\
For 2D pattern generation with arbitrary frequencies, one has to acknowledge that $\alpha$, $\beta$, and $\Xi$ are frequency dependent (\figref{model}).
A simplified approach without efficiency control is to choose global rf power levels that run the AODs close to saturation. 
\figureref[a]{single} shows the measured 2D diffraction efficiency landscape using this approach with the AWG power fraction set to \num{0.9} for both AODs. Contour lines indicate regions in which the AOD system operates above the labeled thresholds.
The maximum diffraction efficiency is \SI{92}{\percent}. While the continuous response allows the user to select local frequency patches with minor variation, the measured landscape reveals a considerable variation even for the central region within the \SI{79}{\percent} contour line. This clearly shows the limits of the unregulated approach.
\begin{figure}[bt!]
    \centering
    \includegraphics[width=\linewidth]{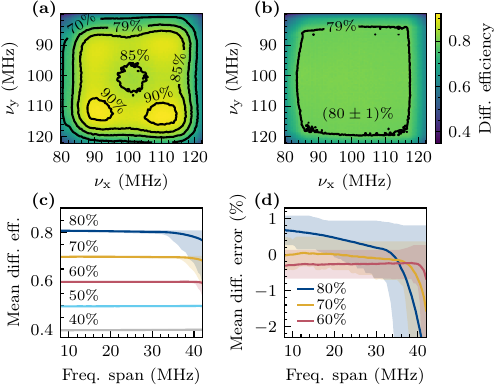}
    \caption{
    Measured 2D diffraction efficiency for single-tone operation at both axes.
    (a) Unregulated operation with fixed AWG power fraction \num{0.9} for both AODs as reference.
    (b) Operation with the control system set to generate a flat response of \SI{80}{\percent} target diffraction efficiency. The color bar on the right holds for (a) and (b).
    (c) The uniformity of the controlled 2D diffraction efficiency is analyzed as a function of the 2D frequency span around $\nu_\textnormal{c}$.
    Mean value and variation (shaded areas) are plotted for \SIrange{40}{80}{\percent} target diffraction efficiency. (d) Mean diffraction error (i.e. the deviation from the target efficiency) and its variation visualizing the precision and accuracy for three selected target values.}
    \label{fig:single}
\end{figure}\\
For many applications, ensuring genuine uniformity of diffraction efficiencies for a given working range of the AOD is crucial. To benchmark the presented control system, we sampled the 2D diffraction efficiency at target levels $\eta^{(\textnormal{xy})}_\textnormal{T}$
between \SI{40}{\percent} and \SI{80}{\percent} in steps of \SI{10}{\percent} over the frequency span of \SI{42}{\mega\hertz} per axis. For all realizations, both single-axis diffraction efficiencies were set equally: $\eta^{(\textnormal{x})}_\textnormal{T}=\eta^{(\textnormal{y})}_\textnormal{T}=(\eta^{(\textnormal{xy})}_\textnormal{T})^{1/2}$.
\figureref[b]{single} gives the result for $\eta^{(\textnormal{xy})}_\textnormal{T}=$ \SI{80}{\percent}.
In comparison to the measurement with fixed rf power, the flatness is strongly improved. Within the \SI{79}{\percent} contour line, \SI{95}{\percent} of the measured efficiencies are in the range \SIrange{79.2}{80.9}{\percent}.
\figureref[c]{single} shows the measured performance for all targeted values $\eta^{(\textnormal{xy})}_\textnormal{T}$. The mean value and the variation of the data for $\eta^{(\textnormal{xy})}_\textnormal{M}$ are plotted as a function of the 2D frequency span around $\nu_\textnormal{c}$. Results confirm that our approach remains linear with flat response independent of the frequencies, as long as the requested efficiency does not exceed the achievable peak efficiency.
The latter effect causes a reduction in the mean value of $\eta^{(\textnormal{xy})}_\textnormal{M}$ and an increased variation for large frequency spans. As displayed in \figref[d]{single}, the edges of the error bands intersect a diffraction efficiency error of \SI{1}{\percent} for a 2D span above \SI{33}{\mega\hertz} for the \SI{80}{\percent} data, above \SI{37}{\mega\hertz} for the \SI{70}{\percent} data, and above \SI{40}{\mega\hertz} for the \SI{60}{\percent} data.
Noting that the apparent precision in this measurement series is better than the specified uncertainty of the laser power meter, we observe two trends for data within the frequency range where the requested efficiency $\eta^{(\textnormal{xy})}_\textnormal{T}$ is still accessible. First, the variation given by the error bands increases with larger frequency spans. Second and more prominently, the curves show a nonzero slope.
We attribute both effects to shortcomings of the model used for the interaction term $\Xi$ (Eq. \eqref{eq:xi-def}) including only up to fourth-order corrections. These effects gain increasing significance for larger deviations from $\xi_\textnormal{x}$ and $\xi_\textnormal{y}$
due to an increased wave vector mismatch (see also \figref[c]{model}).
\subsection{Two-dimensional multibeam efficiency control}
\begin{figure}[bt]
	\includegraphics[width=\linewidth]{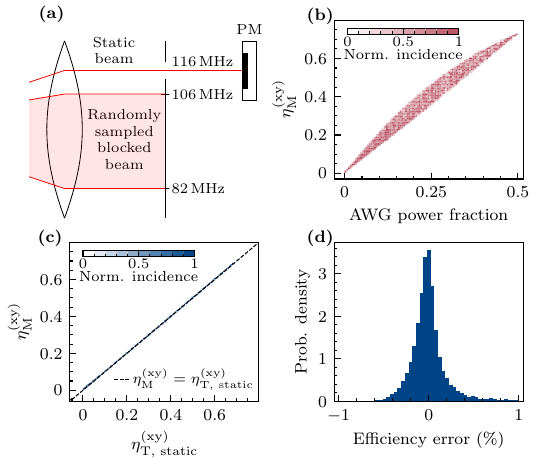}
    \caption{
    Characterization of the crosstalk for dual-tone operation.
    (a) Schematic setup. The y-axis (out of plane) AOD is driven with a constant single-tone signal. For the x-axis AOD, the diffraction efficiency of one tone at the static frequency of \SI{116}{\mega\hertz} is recorded while the power and frequency of a second tone are sampled randomly.
    (b) Measured diffraction efficiency with disabled efficiency control as a function of the static-tone AWG power fraction. Crosstalk-induced spreading of the efficiency is observed. 
    The data are color-coded according to their normalized incidence (\num{2e4} samples, 100 bins per axis). 
    (c) With efficiency control enabled, a linear response as a function of the requested target efficiency is obtained. (d) With efficiency control the residual $2 \sigma$ deviation from the target diffraction efficiency is \SI{\pm 0.5}{\percent}.
    }
	\label{fig:Crosstalk}
\end{figure}
The nonlinear acousto-optic response is additionally evolved for multitone waveforms. All frequency components interfere with the diffraction of each individual tone due to saturation effects. This makes multitone operation highly impractical without efficiency control. To demonstrate the effect, we conducted a measurement of the dual-tone crosstalk on the x-axis AOD while driving the y-axis AOD with a single tone of constant power at $\nu_\textnormal{c}$.
The setup is illustrated in~\figref[a]{Crosstalk}.
The x-axis AOD received a dual-tone signal with one tone remaining static at $\nu_{\text{static}}=\SI{116}{\mega\hertz}$, while the other tone was randomly sampled in the 
range $\nu_{\textnormal{random}} \in[\SI{82}{\mega\hertz},\SI{106}{\mega\hertz}]$. The rf power of both tones was randomly sampled.
The diffracted beams up to \SI{106}{\mega\hertz} were blocked in front of the power meter. Only the static beam was measured.\\
The static y-axis AOD was driven at AWG power fraction \num{0.9}.
Without efficiency control, the power sampling of both x-axis tones was restricted by limiting the combined dual-tone x-axis AWG power fraction to \num{0.5}. This constraint accommodates the crest factor, as a dual-tone signal has a peak power twice as high as a single-tone sine wave of equal mean power (rms).
\figureref[b]{Crosstalk} shows $\eta^{(\textnormal{xy})}_\textnormal{M}$ as a function of the x-axis AWG power fraction of the static tone at \SI{116}{\mega\hertz}.
Without a second tone, the response follows Eq. \eqref{eq:k-mismatch-approx}, as observed for the measurement in \figref[a]{linearity}. For the dual-tone measurement, this corresponds to the samples with no power in the second tone and gives the upper boundary of the data in \figref[b]{Crosstalk}. The interference of the second tone causes a reduction of $\eta^{(\textnormal{xy})}_\textnormal{M}$ for the measured component. The main reduction stems from saturation with increased power values of the second tone. This effect is further enhanced by the frequency dependence of $\beta$ (see \figref[b]{model}).
At \num{0.25} AWG power fraction, the spread in $\eta^{(\textnormal{xy})}_\textnormal{M}$ is \SI{35}{\percent} to \SI{47}{\percent}. The lower boundary arises due to the power constriction of the second tone defined by the maximally allowed dual-tone AWG power fraction, limiting the second tone's power fraction range to \numrange{0}{0.25} at this point.\\
\figureref[c]{Crosstalk} depicts the respective measurement with the efficiency control system in operation. The y-axis AOD was set to $\eta^{(\textnormal{y})}_\textnormal{T}=\num{0.9}$. The dual-tone waveforms for the x-axis AOD were sampled up to a limit of $\eta^{(\textnormal{x})}_{\textnormal{T}}=\num{0.75}$ per tone while constricting samples to combinations that comply with $\eta^{(\textnormal{x})}_\textnormal{T,static}+\eta^{(\textnormal{x})}_\textnormal{T,random}\leq\num{0.75}$ for the combined target efficiency of both tones.
This defines the depicted measurement range of $\eta^{(\textnormal{xy})}_\textnormal{T,static}\in[0,\num{0.675}]$.
The data closely follow the sought relation $\eta^{(\textnormal{xy})}_\textnormal{M} = \eta^{(\textnormal{xy})}_\textnormal{T,static}$ (dashed). \figureref[d]{Crosstalk} shows the deviation from the expected linear response function. The residual error is \SI{\pm 0.5}{\percent} ($2 \sigma$) and below \SI{\pm 1}{\percent} for all samples, demonstrating how effectively our model can compensate for crosstalk between two tones.
\begin{figure}[bt]
    \centering
    \includegraphics[width=\linewidth]{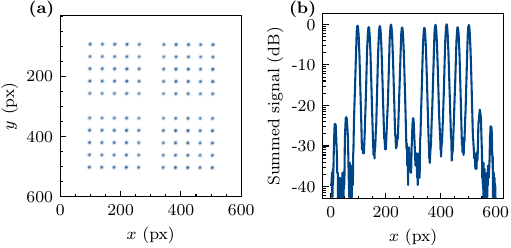}
    \caption{
    Intermodulation analysis of a 2D multispot optical tweezer array. Tweezer positions are defined by a multitone waveform of 11 equally spaced frequencies per axis. Diffraction efficiencies of all tones are equalized while the central rf component is blanked.
    (a) Camera image of the intensity pattern in the back focal plane. The tweezer array covers an area of $\SI{1}{\milli\meter}\times\SI{1}{\milli\meter}$.
    (b) Camera signal summed along the y axis to reveal any intermodulation-induced signal. The gap shows a third-order intermodulation peak at \SI{-23}{dB}.
    }
    \label{fig:LightIntermodulation}
\end{figure}\\
Multibeam efficiency control is applied to generate the 2D beam patterns of \figref[b]{f0} and \figref[c]{f0} in real-time mode.
A quantitative analysis of the efficiency control of large 2D laser spot arrays is performed for the pattern of \figref[b]{f0}. For this intensity-equalized 50$\times$50 array, the measured integral diffraction efficiency of \num{0.495 \pm 0.023} 
agrees well with the targeted efficiency of 0.5. When analyzing the homogeneity of all 2500 laser spots, we observe a relative rms spread of \SI{5.3}{\percent}. This performance is comparable to significantly more involved methods that employ extensive calibration of a static array implementation \cite{CooperRoy2018}.
\subsection{Intermodulation distortion}
\label{sec:intermodulation}
A major source of disturbances in multitone pattern generation is intermodulation. Although control of intermodulation is outside the scope of the applied model, we investigated its effect since we attribute part of the intensity variation in \figref[b]{f0} to intermodulation.
Due to nonlinearities in signal processing, including amplification and signal transduction to the AOD's crystal, spurious signals are created. A poorly optimized crest factor strongly favors intermodulation.
The leading order is found at the sum and difference frequencies of pairs of programmed tones for the respective axis which are usually outside the range of operation. In next order, third-order intermodulation gives rise to frequency components that lie within the frequency range of the original waveform.\\
For a multitone intermodulation analysis, we recorded camera images of 2D tweezer patterns that were deliberately chosen to be prone to intermodulation by using grids of equally spaced frequencies and setting the amplitudes for the central frequency components (i.e., the target power for the central row and column) to zero. Intermodulation of neighboring tones could create a spurious signal at these exact frequencies and corresponding beam positions. 
\figureref[a]{LightIntermodulation} shows a realization with $11$ tones in both dimensions spaced by \SI{0.75}{\mega\hertz}, creating four clusters of 5$\times$5 equidistant tweezers that are separated by a central gap of \SI{1.5}{\mega \hertz}. An integral target diffraction efficiency of 0.3 was set for the grid and \num{0.245 \pm 0.023} was measured.
Although we attribute this reduction of efficiency to intermodulation, no spurious spots are immediately visible on the 2D camera image of \figref[a]{LightIntermodulation}. To quantify the magnitude the intermodulation in the light field, we integrated the camera signal along the y axis as shown in \figref[b]{LightIntermodulation}. For the blanked frequency at pixel 300, an optical signal with an intensity reduction of \SI{23}{dB} is visible with similar peaks left and right of the target beam pattern. The uniformity of the spectrum is also affected, with a relative rms spread of \SI{9.8}{\percent}.
These measurements highlight the importance of taking into account the characteristics of the specific device for high-fidelity multitone operation. 
Crest factor optimization with direct methods is particularly suitable for multitone patterns with fixed frequency spacing, but time-varying frequency spacings may result in reduced suppression of intermodulation which has to be closely monitored and presumably mitigated by predistortion methods \cite{Biglieri1988, Morgan2006}. Since predistortion and advanced phase tuning for crest factor reduction strongly depend on the acousto-optic device as well as the application, their optimization is beyond the scope of this work.
\section{Conclusion}
Building on the theoretical framework of the acousto-optic response that is described by coupled-wave theory, we deduced a resource-efficient model for the computation of the diffraction efficiencies of multitone signals in AODs. The description enfolds 2D pattern generation by incorporating geometric effects that arise from operating two mutually perpendicular AODs in series.
Model-parameter determination only requires measurements of the laser power after 2D deflection with a single-tone signal at each axis. This allows the straightforward integration in existing setups and optical instruments since model calibration can be realized by addition of a simple beam splitter and a power meter.\\
On this basis, we developed an open-loop control system that ensures strict intensity matching for user-programmable random-access scanning applications with acousto-optically deflected laser beams. Our tailored software implementation inverts the model in real time on commodity hardware to determine the required rf power per frequency component to achieve a specific diffraction efficiency target pattern, enabling on-the-fly control of intensity profiles for arbitrary spatiotemporal multibeam trajectories.
In the configuration presented, it is capable of streaming tweezer arrays with up to 2500 spots in continuous mode. Waveforms are generated with adequate time resolution for controlling common AODs at their full acousto-optical bandwidth.\\
The performance of the approach presented was validated in a 2D AOD setup. Linear control over diffraction efficiencies in single-tone and multitone operation was achieved. Residual errors were within the measurement uncertainty. Intermodulation was found to be sufficiently suppressed by direct methods for crest factor optimization that are feasible for real-time operation.
The presented real-time system for reactive 2D multibeam laser scanning will benefit abundant applications that employ optical forces by optical tweezers or precise deposition of light energy with exquisite spatiotemporal resolution. In our work, we focus on quantum science with arrays of atomic qubits, where laser addressing of individual qubits is essential at every stage of operation.
Parallelized qubit transport will strongly improve the preparation of defect-free many-body systems~\cite{Tian2023,Wang2023,GuoX2025,Lin2024,Afiouni2025,Dadpour2025}, their continuous operation~\cite{Singh2022a,Pause2023,Norcia2024,Gyger2024,Yan2025,Chiu2025}, as well as increasing data rates in sensing applications~\cite{Schaffner2024}.
Moreover, time-critical parallelized optical control at the individual qubit level becomes exceedingly momentous in quantum information processing. The application of quantum algorithms and error correction protocols~\cite{Bluvstein2024,Reichardt2024,Bedalov2024,Rodriguez2024} necessitates local qubit addressing for coherent quantum-state control, mid-circuit measurements, conditional quantum operations, and memory transfer in zoned architectures~\cite{Tan2024}. High-fidelity rendering of programmed intensities, as demonstrated in this work, is vital. Deviations directly impact the quantum-state phases of the qubits, scrambling quantum information.
For large-scale implementations~\cite{Pause2024,Pichard2024,Manetsch2025,Lin2024}, high-bandwidth multibeam steering is of immediate relevance for future progress.\\
\begin{acknowledgments}
We thank S.~de~Léséleuc for insightful discussions.
We acknowledge financial support by the Federal Ministry of Education and Research (BMBF) [Grant No. 13N15981], by the Federal Ministry of Research, Technology and Space (BMFTR) [Grant No. 13N17366], and by the Deutsche Forschungsgemeinschaft (DFG -- German Research Foundation) [Grant No. BI 647/6-1 and BI 647/6-2, Priority Program SPP 1929 (GiRyd)].
\end{acknowledgments}
\bibliographystyle{jabbrv_apsrev4-1}
\bibliography{Multitone}

\end{document}